\documentclass[floafix,prd,showpacs,showkeys,preprintnumbers,nofootinbib,superscriptaddress,11pt]{revtex4-1}
\usepackage[utf8]{inputenc}
\usepackage[sort&compress]{natbib}
\usepackage{ulem}
\usepackage{bm}
\usepackage{times}
\usepackage{amssymb,amsbsy,amsmath,amsfonts}
\usepackage{graphicx}
\usepackage{epstopdf}
\usepackage{float}
\usepackage{color}
\usepackage{morefloats}
\usepackage{rotating}
\usepackage{srcltx}
\usepackage{slashed}
\usepackage{subfigure}
\usepackage{multirow}
\usepackage{verbatim}
\usepackage{hyperref}
\usepackage{lineno}
\usepackage{overpic}
\usepackage{booktabs}
\usepackage{tabularx}

\begin{document}

\title{ $B$-meson decays to vector charmoniumlike states and a $K$  meson: The role of final-state interactions }

\author{Qi-Wei Yuan}
\affiliation{
Frontiers Science Center for Rare Isotopes, Lanzhou University,
Lanzhou 730000, China}
\affiliation{ School of Nuclear Science and Technology, Lanzhou University, Lanzhou 730000, China}

\author{Qi Wu}
\email[Corresponding author:]{wuqi@htu.edu.cn}
\affiliation{
Institute of Particle and Nuclear Physics, Henan Normal University, Xinxiang 453007, China}

\author{Ming-Zhu Liu}
\email[Corresponding author:]{liumz@lzu.edu.cn}
\affiliation{
Frontiers Science Center for Rare Isotopes, Lanzhou University,
Lanzhou 730000, China}
\affiliation{ School of Nuclear Science and Technology, Lanzhou University, Lanzhou 730000, China}

\begin{abstract}
A series of vector charmonium(like) states, accompanied by a $K$ meson, have been observed in the decays of $B$ meson. These processes are color-suppressed at the quark level, as inferred from topological diagram analysis.   In this work, we calculate the branching fractions of the  decays   $B \to \psi K$, where $\psi$ denotes the charmonium(like) states $\psi(1S)$, $\psi(2S)$, $\psi(4040)$, $\psi(3770)$, and $\psi(4160)$. Our analysis incorporates both short-distance (naive factorization approach) and long-distance (final-state interactions) contributions.    Within reasonable parameters, our results align with experimental data except for the $ \psi(4160)$, suggesting its possible exotic nature.     Furthermore, we find that  long-distance contributions dominate these decay processes, highlighting the crucial role of final-state interactions in  the productions of charmonium(like) states in $B$ decays.                          

\end{abstract}


\maketitle

\section{Introduction}

With the large data samples collected by $B$ factories and LHCb experiments,  
 $B$ meson decays have become  an ideal  platform for precisely testing the Standard Model and searching for  new physics signals~\cite{LHCb:2008vvz,Belle-II:2018jsg}. Surprisingly,  many new hadron states beyond the conventional mesons and baryons in the quark model, also known as exotic states,  have been  observed in  $b$-flavored hadron decays,  which provide the opportunity to deeply study the nonperturbative strong interaction as well as explore the existence of hadrons with unconventional configurations~\cite{Brambilla:2010cs}. Over the past two decades, revealing the nature of these exotic states has been a primary focus for both experimentalists and theorists. However, their underlying structure remains highly debated~\cite{Chen:2016qju,Lebed:2016hpi,Oset:2016lyh,Esposito:2016noz,Dong:2017gaw,Guo:2017jvc,Olsen:2017bmm,Ali:2017jda,Karliner:2017qhf,Guo:2019twa,Brambilla:2019esw,Meng:2022ozq,Liu:2024uxn,Wang:2025sic}. Among them, the charmoniumlike states as  one
of the major categories  motivate intensive discussions on their internal structure and properties~\cite{Brambilla:2019esw}.

The recent studies show that the charmoniumlike states have strong couplings to a pair of charmed mesons. The unquenched quark model has been utilized to analyze the mass spectra of charmoniumlike states, demonstrating that the inclusion of charmed meson pair effects significantly reduces the discrepancies between theoretical predictions and experimental measurements~\cite{Barnes:2007xu,Ortega:2012rs,Kanwal:2022ani,Deng:2023mza,Man:2024mvl,Pan:2024xec}. On the other hand,   the spectra  of these  states can also be described at the hadron level with  the hadron-hadron interaction dressed by the presence of bare states~\cite{Yamaguchi:2019vea,Ji:2022vdj,Song:2023pdq,Kinugawa:2023fbf,Shi:2024llv}. Moreover, the strong decays of charmoniumlike states have shown that charmed meson pairs play a crucial role in explaining the experimental observations~\cite{Meng:2007cx,Zhang:2009kr,Guo:2010ak,Li:2013zcr}. It is evident that hadronic molecule components and bare components are intricately mixed in charmoniumlike states, and these two components may compete in the formation of such states. For some states near the mass thresholds of charmed meson pairs such as $X(3872)$~\cite{Belle:2003nnu}, the hadronic molecule components account for a significant proportion~\cite{Song:2022yvz,Kinugawa:2022fzn}, which are  even regarded as pure hadronic molecules.


It is natural to investigate  the internal structure  of the charmoniumlike states through $B$ meson decays.     However, calculating their branching fractions at the quark level is challenging due to the intricate nonperturbative effects and the ambiguous internal structures of charmoniumlike states~\cite{Song:2003yc}.  The production of the $X(3872)$ in $B$ meson decay has been garnered significant interest.  In Ref.~\cite{Meng:2005er}, Meng \textit{et al.}, adopted the QCD factorization approach well explaining the absolute branching fraction of the decay $B \to X(3872) K$, but failing to explain‌ the ratio of  $\mathcal{B}[B^+ \to X(3872) K^+]/\mathcal{B}[B^0 \to X(3872) K^0]$,  where $X(3872)$ is assumed to be a $\chi_{c1}(2P)$ state.  On the other hand, by assuming $X(3872)$ to be a  $\bar{D}D^*$ molecule, we utilized the final state interactions (FSIs) approach to successfully describe both the absolute branching fraction and their ratio~\cite{Wu:2023rrp}, suggesting  that $X(3872)$ contains a significant molecular component, consistent with experimental analyses~\cite{LHCb:2020xds,BESIII:2023hml}. Furthermore, our results indicate that the long-distance contributions   dominate the decays $B \to X(3872) K$, similar to those of other charmoniumlike states~\cite{Xie:2022lyw,Liu:2024hba,Yu:2024mol}.        The FSIs approach has already been  extensively utilized to address the nonperturbative effects inherent in the decays of heavy-flavored hadrons~\cite{Yu:2017zst,Hsiao:2019ait,Duan:2024zjv}.       In this work, we employ the FSIs approach to study  the production of vector charmonium(like)   states  in $B$ decays.

\begin{table}[ttt]
\caption{Branching fractions~( in units of $10^{-4}$) for the decays of  $B$ mesons  into vector charmonium(like) states accompanied by a kaon meson~\cite{ParticleDataGroup:2024cfk}. \label{Experimentbranching}}
\begin{tabular}{c|ccccccccc}
  \hline\hline
   Decay modes &  ~~~$\to \psi(1S)$~~~  & ~~~$\to \psi(2S)$~~~  &   ~~~$\to \psi(4040)$~~~  &  ~~~$\to \psi(3770)$~~~   & ~~~$\to \psi(4160)$~~~ &   \\ \hline
       Experiments &  $10.20\pm0.19$  & $6.24\pm0.21$   &  $16.0\pm 5.0$  &   $4.3\pm1.1$   & $5.1\pm 2.7$  &    \\
 \hline \hline
\end{tabular}
\end{table}

It should be noted that  the number of experimental observations of the vector charmonium(like) states is the largest among all charmonium(like) states~\cite{ParticleDataGroup:2024cfk}, and this trend also holds true for their observations in $B$  decays. In Table~\ref{Experimentbranching},  we list  the observed  experimental branching fractions of the $B$ meson decaying into the vector charmonium(like) states and a $K$ meson.  The absolute branching fraction of the decay $ B \to \psi(4415) K$ is absent although $\psi(4415)$ is observed in the decay $B\to D\bar{D}K$~\cite{ParticleDataGroup:2024cfk}. 
According to the potential model~\cite{Godfrey:1985xj,Barnes:2005pb,Li:2009zu,Deng:2023mza},  the charmoniumlike states $\psi(4040)$, $\psi(3770)$, and $\psi(4160)$ are assigned as the $\psi(3S)$,  $\psi(1D)$, and  $\psi(2D)$ states, where the excited bare components account for large proportion. By analyzing the experimental data~\cite{Zhou:2023yjv},  $\psi(4040)$ and  $\psi(3770)$ are assigned as  $\psi(3S)$ and   $\psi(1D)$ states, while  $\psi(4160)$ and $\psi(4230)$ are assigned as  the same state $\psi(2D)$\footnote{Unlike  $\psi(3770)$, the pole positions of  $\psi(4040)$  and $\psi(4160)$ exist large uncertainty~\cite{Husken:2024hmi,Peng:2024blp}.}.    Up to now,  a unified  framework to interpret the branching fractions of the $B$ meson decaying into the vector charmonium(like) states and a $K$ meson, as listed in   Table~\ref{Experimentbranching},   is still
missing, which  is the primary objective of the present work.

This work is organized as follows. In Sec.~II,  we begin by briefly introducing the tree-level Feynman diagrams and triangle Feynman diagrams, which account for the short-distance and long-distance contributions, respectively, in the decays of the $B$ meson into vector charmonium(like) states and a $K$ meson. We also present the effective Lagrangian approach in this section.   The numerical results and discussions are provided  in Sec.~III. Finally, a summary is given in the last section.


\section{ THEORETICAL FORMALISM }
\label{sec:Sec2}

\begin{figure}[ttt]
\begin{center}
\begin{overpic}[scale=0.98]{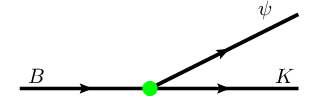}
\end{overpic}
  \caption{Tree diagram accounting  for short-distance contributions to the decays $B \rightarrow  \psi K $ [$\psi(1S)$, $\psi(2S)$, $\psi(4040)$, $\psi(3770)$, and $\psi(4160)$]. }
  \label{Fig:tree}
\end{center}
\end{figure}

\begin{figure}[ttt]
\begin{center}
\begin{tabular}{ccc}
\subfigure[][]
{
\begin{minipage}[t]{0.32\linewidth}
\begin{center}
\begin{overpic}[scale=1]{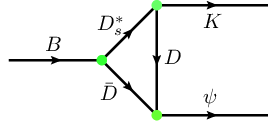}
\end{overpic}
\end{center}
\end{minipage}
}
\subfigure[][]
{
\begin{minipage}[t]{0.32\linewidth}
\begin{center}
\begin{overpic}[scale=1]{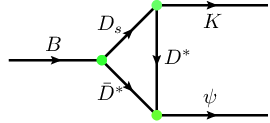}
 \end{overpic}
\end{center}
\end{minipage}
}
\subfigure[][]
{
\begin{minipage}[t]{0.32\linewidth}
\begin{center}
\begin{overpic}[scale=1]{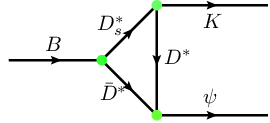}
\end{overpic}
\end{center}
\end{minipage}
}
\\
\subfigure[][]
{
\begin{minipage}[t]{0.32\linewidth}
\begin{center}
\begin{overpic}[scale=1]{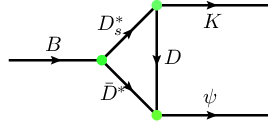}
\end{overpic}
\end{center}
\end{minipage}
}
\subfigure[][]
{
\begin{minipage}[t]{0.32\linewidth}
\begin{center}
\begin{overpic}[scale=1]{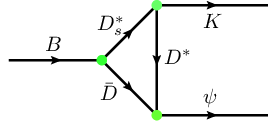}
\end{overpic}
\end{center}
\end{minipage}
}
\subfigure[][]
{
\begin{minipage}[t]{0.32\linewidth}
\begin{center}
\begin{overpic}[scale=1]{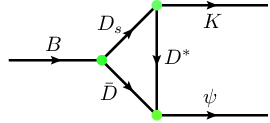}
\end{overpic}
\end{center}
\end{minipage}
}
\\ 
\end{tabular}
\caption{Triangle diagrams accounting for long-distance contributions to the decays $B \rightarrow   \psi K$ [$\psi(1S)$, $\psi(2S)$, $\psi(4040)$, $\psi(3770)$, and $\psi(4160)$]. All diagrams contribute to the productions of \(\psi(1S)\), \(\psi(2S)\), and \(\psi(4040)\) in $B$ decays. Only the diagram (a)  contributes to the production of \(\psi(3770)\) in $B$ decay, and the diagrams (b)-(f) contribute to the production of \(\psi(4160)\) in $B$ decay.   }\label{Fig:loop2}
\end{center}
\end{figure}

 The decays of the $B$ meson into the vector charmonium(like) states and a $K$ meson (denoted by $B \to \psi K$ ) proceed via the quark-level transition  $b \to c \bar{c}s$~\cite{Chen:2021erj}.
From the analysis of the  topological diagrams, the decays  $B \to \psi K$ belong to  $C$-type diagrams, which are  color suppressed~\cite{Mantry:2003uz,Leibovich:2003tw}. As indicated in Ref.~\cite{Colangelo:2002mj}, the nonfactorizable effects responsible for the long-distance contributions are sizable in  these decays.    Therefore, following Refs.~\cite{Cheng:2004ru,Duan:2024zjv},  we consider both the short-distance and long-distance contributions to the decays  $B \to \psi K$\footnote { As indicated in Ref.~\cite{Chen:2005ht},  the  $B \to \psi K$ decays are classified as the factorization and nonfactorization contributions, which are described by naive factorization approach and   perturbative QCD approach, respectively. However, the PQCD approach would be challenged for the higher excited charmonium states due to unknown hadron distribution amplitudes.  In contrast to classification of the  heavy-hadron weak decays as the factorization and nonfactorization contribution,  their amplitudes can be derived  by $SU(3)$-flavor symmetry~\cite{Roy:2019cky,He:2015fwa} and  factorization assisted
topological amplitude approach~\cite{Zhou:2016jkv},  which are heavily relied on the experimental data.     }.    The short-distance contributions are depicted by tree diagrams  as shown in Fig.~\ref{Fig:tree}, while the long-distance contributions are represented by triangle diagrams~\cite{Colangelo:2003sa,Wang:2008hq,Xu:2016kbn} as shown in Fig.~\ref{Fig:loop2}.

\begin{table}[ttt]
    \centering
    \caption{ $f_\psi$ decay constants of the vector charmonium(like) states.}
    \label{decayconstantvector}
\begin{tabular}{c|ccccccccc}
  \hline\hline
   States &  ~~~$ \psi(1S)$~~~  & ~~~$ \psi(2S)$~~~  &   ~~~$ \psi(4040)$~~~  &  ~~~$ \psi(3770)$~~~   & ~~~$ \psi(4160)$~~~ &   \\ \hline
       Decay Constants(MeV) &  415.49  & 294.03   &  186.81  &   99.67   & 141.83 &    \\
 \hline \hline
\end{tabular}
\end{table}

With the factorization ansatz,  the tree diagram of the decays $B \to \psi K$  can be expressed as the product of two current matrix elements
\begin{eqnarray}\label{BtoX387245}
\mathcal{A}\left(B \to \psi K\right)&=&\frac{G_{F}}{\sqrt{2}} V_{cb}V_{cs} a_{2}\left\langle \psi|(c\bar{c})_{V-A}| 0\right\rangle\left\langle K |(s \bar{b})_{V-A}| B\right\rangle,
\end{eqnarray}
with $(q\bar{q})_{V-A}$ standing for $q\gamma_{\mu}(1-\gamma_5)\bar{q}$.  
Since the effective Wilson coefficient   $a_2$ depends on the renormalization energy scale,   we adopt the value of  $a_2$ at the mass of bottom quark, i.e., $a_2=0.103$~\cite{Lu:2009cm}.   In this work, we take the values of  $G_F = 1.166 \times 10^{-5}~{\rm GeV}^{-2}$, $V_{cb}=0.041$,   $V_{cs}=0.987$.   
The current matrix element for the $\psi$ meson  created from the vacuum  is parametrized as~\cite{Cheng:2003sm,Verma:2011yw}  
\begin{eqnarray}
\left\langle \psi|(c\bar{c})_{V-A}| 0\right\rangle &=& f_{\psi} m_{\psi} \varepsilon_{\mu}^*, \label{Eq:0Y}
\end{eqnarray}
where $f_{\psi}$ is the decay constant of a $\psi$ meson. Given the  electronic width of a $\psi$ meson, one can derive its decay constant via 
 $ f_\psi=\sqrt{\frac{3m_\psi\Gamma^{\psi \rightarrow e+e^-}}{4\pi\alpha^2e_q^2}}$~\cite{Shi:2023ntq,Liu:2024hba}, 
where $\alpha$ is the  fine-structure constant.  The decay constants of $\psi(1S)$, $\psi(2S)$, $\psi(4040)$, $\psi(3770)$, and $\psi(4160)$ mesons are presented in Table~\ref{decayconstantvector}.  
The current matrix element  of  $\left\langle K| (s\bar{b})_{V-A}| B\right\rangle$ describing the hadronic transitions
is parametrized by the following form factors:
\begin{eqnarray}
\left\langle K| (s\bar{b})_{V-A}| B\right\rangle =\left[P^{\mu}-\frac{m_{B}^2-m_{K}^2}{q^2}q^{\mu}\right] F_{1}(q^2)+\frac{m_{B}^2-m_{K}^2}{q^2}q^{\mu} F_{0}(q^2),\label{Eq:BK}
\end{eqnarray}
with $F_0$ and $F_1$ being  the form factors for $B(k_0)\to K(p_1)$,  where  $P=k_0+p_1$ and $q=k_0-p_1$.

With the above effective Lagrangian, we  obtain the  amplitude of the decays $B(k_0) \to \psi(q)K(p_1)$ in Fig.~\ref{Fig:tree}  as   
\begin{eqnarray}\label{Ds-KK}
\mathcal{A}\left(B\to\psi K\right)&=&\frac{G_{F}}{\sqrt{2}} V_{cb}V_{cs} a_{2} f_{\psi} m_{\psi} (k_0+p_1)\cdot \varepsilon(p_2) F_{1}(q^2).
\end{eqnarray}

For the weak interaction  vertices of triangle diagrams in Fig.~\ref{Fig:loop2}, the  decay amplitudes of $B(k_0)\to  D_{s}^{(\ast)}(q_1) \bar{D}^{(\ast)}(q_2)$  can be expressed as the products of two current matrix elements~\cite{Ali:1998eb,Qin:2013tje}
\begin{eqnarray}\label{Ds-KK}
\mathcal{A}\left(B\to D_{s} \bar{D}^{\ast}\right)&=&\frac{G_{F}}{\sqrt{2}} V_{cb}V_{cs} a_{1}\left\langle D_{s}|(s\bar{c})_{V-A}| 0\right\rangle\left\langle \bar{D}^{\ast }|(c \bar{b})_{V-A}| B\right\rangle , \\
\mathcal{A}\left(B\to D_{s}\bar{D}\right)&=&\frac{G_{F}}{\sqrt{2}} V_{cb}V_{cs} a_{1} \left\langle D_{s}|(s\bar{c})_{V-A}| 0\right\rangle\left\langle \bar{D}^{ }|(c \bar{b})_{V-A}| B\right\rangle , \\
\mathcal{A}\left(B\to D_{s}^{\ast} \bar{D}\right)&=&\frac{G_{F}}{\sqrt{2}} V_{cb}V_{cs} a_{1} \left\langle D_{s}^{\ast}|(s\bar{c})_{V-A}| 0\right\rangle\left\langle \bar{D}|(c \bar{b})_{V-A}| B\right\rangle , \\ 
\mathcal{A}\left(B\to D_{s}^{\ast} \bar{D}^{\ast}\right)&=&\frac{G_{F}}{\sqrt{2}} V_{cb}V_{cs} a_{1} \left\langle D_{s}^{\ast}|(s\bar{c})_{V-A}| 0\right\rangle\left\langle \bar{D}^{\ast }|(c \bar{b})_{V-A}| B\right\rangle.
\end{eqnarray}
Similarly, we adopt the effective Wilson coefficient $a_1=1.03$~\cite{Lu:2009cm}.  The former current matrix elements  are written as
\begin{eqnarray}
\langle D_s(p_1)|(s\bar{c})_{V-A}|0\rangle&=&i f_{D_s} p_{1\mu},   \\
\langle D_{s}^*(p_1,\epsilon)|(s\bar{c})_{V-A}|0\rangle&=&f_{D_{s}^*}  m_{D_{s}^*}\epsilon_\mu^*,
\label{Eq:ME1}
\end{eqnarray}
where   the  decay constants of $D_{s}$ and $D_{s}^*$ are adopted  as  $f_{D_{s}} = 250$ MeV and $f_{D_{s}^{\ast}}=272$~MeV~\cite{ParticleDataGroup:2022pth,Verma:2011yw,FlavourLatticeAveragingGroup:2019iem,Li:2017mlw,Li:2012cfa}.
The  matrix elements of hadron transitions  $B(k_0) \to \bar{D}^{(*)}(q_2)$  are parametrized by following  form factors~\cite{Cheng:2003sm,Verma:2011yw}
\begin{eqnarray}
&&\left\langle \bar{D}^{\ast}|(c\bar{b})_{V-A}| B\right\rangle=i\epsilon_{\alpha}^{*}\left\{-g^{\mu \alpha} (m_{\bar{D}^{\ast}}+m_{B}) A_{1}\left(q^{2}\right)+P^{\mu} P^{\alpha} \frac{A_{2}\left(q^{2}\right)}{m_{\bar{D}^{\ast}}+m_{B}}\right. \\  \nonumber
&&+i \varepsilon^{\mu \alpha \beta \gamma}P_\beta q_\gamma \left.\frac{V\left(q^{2}\right)}{m_{\bar{D}^{\ast}}+m_{B}} +q^{\mu} P^{\alpha} \left[\frac{m_{\bar{D}^{\ast}}+m_{B}}{q^{2}}A_{1}\left(q^{2}\right)-\frac{m_{B}-m_{\bar{D}^{\ast}}}{q^{2}}A_{2}\left(q^{2}\right)-\frac{2m_{\bar{D}^{\ast}}}{q^{2}}A_{0}\left(q^{2}\right)\right]\right\},
\\ 
&&\left\langle \bar{D}|(c \bar{b} )_{V-A}| B\right\rangle =\left[P^{\mu}-\frac{m_{B}^2-m_{\bar{D}}^2}{q^2}q^{\mu}\right] F_{1}(q^2)+\frac{m_{B}^2-m_{\bar{D}}^2}{q^2}q^{\mu} F_{0}(q^2),
\end{eqnarray}
with $A_{1,2,3}$, $V$  being  the form factors,  where  $q= k_{0}- q_2$ and $P = k_0+q_2$. 

  { 
 The form factors are parametrized by   three-parameter form 
\begin{eqnarray}
F(q^2)=\frac{F(0)}{1-a\zeta+b\zeta^2},\label{Eq:A1}
\end{eqnarray}
with  $\zeta=q^2/m^2_{B}$, where the values of parameters $F(0)$, $a$, and $b$ are extracted by the covariant light-front quark model~\cite{Verma:2011yw}, listing in Table~\ref{BtoDformfactor}. The $z$ series expansion provides a more model-independent approach compared to this three-parameter form, such as the Bourrely-Caprini-Lellouch (BCL) parametrization~\cite{Bourrely:2008za} and the Boyd-Grinstein-Lebed (BGL) parametrization~\cite{Boyd:1997kz}.  The BCL and BGL are responsible for the final states of light hadrons and heavy hadrons, respectively. Recently, using an improve QCD approach, the form factors of $B \to K$ and  $B \to D^{(\ast)}$  are precisely calculated by  BCL parametrization~\cite{Cui:2022zwm} and BGL parametrization~\cite{Cui:2023jiw}, respectively. In this work,  the form factors  only take the value at  $q^2=M^2$, and then they  reduce to a set of constants. Therefore, the form factors do not  contribute to the integral of triangle diagram. We find that the values of the form factors at $q^2=M^2$ in 
 Ref.~\cite{Verma:2011yw} and Refs.~\cite{Cui:2022zwm,Cui:2023jiw} have differences at the level of $10\%$.         }

 \begin{table}[ttt]
 \centering
 \caption{Values of  $F(0)$, $a$, and  $b$ in the $B \rightarrow D^{(*)}$ and $B \rightarrow K$ transition form factors~\cite{Verma:2011yw}. \label{BtoDformfactor} }
 \begin{tabular}{ccccccc|ccccccc}
 \hline\hline  
   &   $V$~~~ & $A_0$~~~ &  $A_1$~~~ & $A_2$~~~   & $F_0$~~~   & $F_1$~~~   &~~~~~~  & $F_0$~~~   & $F_1$~~~  \\
 \hline
 $F(0)^{B\to D^{(*)}}$&  0.77~~~  & 0.68~~~ & 0.65~~~ & 0.61~~~   & 0.67~~~ & 0.67~~~ & $F(0)^{B\to K}$   & 0.34~~~ & 0.34~~~ \\
 $a^{B\to D^{(*)}}$  & 1.25~~~ &  1.21~~~  & 0.60~~~ & 1.12~~  & 0.63~~~ & 1.22~~~ & $a^{B\to K}$ & 0.78~~~ & 1.60~~~\\
 $b^{B\to D^{(*)}}$ & 0.38~~~  & 0.36~~~  & 0.00~~~ & 0.31~~~  & -0.01~~ & 0.36~~~ & $b^{B\to K}$ & 0.05~~~ & 0.73~~~ \\ 
\hline\hline
 \end{tabular}
 \end{table}

 With above Lagrangians, the  weak decay amplitudes of $B(k_0)\to D_{s}^{(\ast)}(q_1)\bar{D}^{(\ast)}(q_2)$ have the following form
\begin{align}\label{am3}
\mathcal{A}(B\to D_{s}\bar{D}^{\ast})&= -\frac{G_{F}}{\sqrt{2}}V_{cb}V_{cs} a_{1} f_{D_{s}}\{-q_{1}\cdot \varepsilon(q_{2})(m_{\bar{D}^{\ast}}+m_{B}) A_{1}\left(q_{1}^{2}\right)   \\ \nonumber 
&+(k_{0}+q_{2}) \cdot \varepsilon(q_{2}) q_{1}\cdot (k_{0}+q_{2}) \frac{A_{2}\left(q_{1}^{2}\right)}{m_{\bar{D}^{\ast}}+m_{B}} +(k_{0}+q_{2}) \cdot \varepsilon(q_{2}) \\ \nonumber       & 
[(m_{\bar{D}^{\ast}}+m_{B})A_{1}(q_{1}^2) -(m_{B}-m_{\bar{D}^{\ast}})A_{2}(q_1^2) -2m_{\bar{D}^{\ast}} A_{0}(q_{1}^2)]  \} , \\  
\mathcal{A}(B\to D_{s}\bar{D})&=i\frac{G_{F}}{\sqrt{2}}V_{cb}V_{cs} a_{1}f_{D_{s}}(m_{B}^2-m_D^2)F_{0}(q_{1}^2),
\\  
\mathcal{A}(B\to D_{s}^{\ast}\bar{D})&= \frac{G_{F}}{\sqrt{2}}V_{cb}V_{cs} a_{1}m_{D_{s}^{\ast}}f_{D_{s}^{\ast}}(k_{0}+q_{2})\cdot \varepsilon(q_{1})F_{1}(q_{1}^2), \\ 
 \mathcal{A}(B\to D_{s}^{\ast}\bar{D}^{\ast})&= i\frac{G_{F}}{\sqrt{2}}V_{cb}V_{cs} a_{1}m_{D_{s}^{\ast}}f_{D_{s}^{\ast}}\varepsilon_{\mu}(q_1)\left[-g^{\mu \alpha} (m_{\bar{D}^{\ast}}+m_{B}) A_{1}\left(q_1^{2}\right)\right.  \\ \nonumber &+ \left. (k_0+q_2)^{\mu} (k_0+q_2)^{\alpha} \frac{A_{2}\left(q_1^{2}\right)}{m_{\bar{D}^{\ast}}+m_{B}}  
+i \varepsilon^{\mu \alpha \beta \gamma}(k_0+q_2)_\beta q_{1\gamma} \frac{V\left(q_1^{2}\right)}{m_{\bar{D}^{\ast}}+m_{B}}\right]\varepsilon_{\alpha}(q_{2}).
\end{align}

The Lagrangians  describing the interactions between charmed mesons and the kaon~\cite{Oh:2000qr}
\begin{eqnarray}
\label{dddds}
\mathcal{L}_{D_{s}^{\ast} D K}&=& -i g_{D_{s}^{\ast} D K} ( D \partial^{\mu} K \bar{D}_{s\mu}^{\ast}-  D_{s\mu}^{\ast}\partial^{\mu} K \bar{D}), \\  
\mathcal{L}_{D_{s} D^{\ast} K}&=& -i g_{D_{s} D^{\ast} K} (D_{s} \partial^{\mu} K \bar{D}^{\ast}_{\mu}-D_{\mu}^* \partial^{\mu} K  \bar{D}_{s}), \\  
\label{dddds1}
\mathcal{L}_{D_{s}^{\ast} D^{\ast} K}&=& - g_{D_{s}^{\ast} D^{\ast} K} \varepsilon_{\mu\nu\alpha\beta} \partial^{\mu}D_{s}^{\ast\nu} {\partial}^{\alpha} \bar{D}^{\ast\beta} K , 
\end{eqnarray}
where the couplings are determined as   $g_{ {D}_{s}^*D K}=g_{ {D}_{s}D^* K}=14.00\pm1.73$ and  $g_{ {D}_{s}^*D^* K}=8.63\pm0.99$~GeV$^{-1}$  via SU(3)-flavor symmetry~\cite{Bracco:2011pg}.  

\begin{table}[ttt]
\caption{ Partial  widths of the charmoniumlike states decaying into a pair of charmed mesons. \label{decaywidths}}
\begin{tabular}{c|ccccccccc}
  \hline\hline
    &  $\psi(4040)_{I}$~\cite{ParticleDataGroup:2024cfk}  &  $\psi(4040)_{II}$~\cite{Husken:2024hmi}   &    $\psi(4160)_{I}$~\cite{ParticleDataGroup:2024cfk} &  $\psi(4160)_{II}$~\cite{Peng:2024blp} &    $\psi(3770)$~\cite{ParticleDataGroup:2024cfk} &   &   \\ \hline
  Total & $84.5\pm12.3$&  $130\pm30$  &    $69\pm10$   &    $104.83\pm23.71$   & $27.5\pm0.9$ &   &   \\
    $\bar{D}D$ & $14.28\pm2.08$   &  $40.81\pm9.42$ & $...$   & $27.18\pm6.15$ & $25.58\pm0.84$  &       \\
        $\bar{D}^*D$ & $29.75\pm4.33$ & $37.79\pm8.72$  &  $8.75\pm1.26$ &  $14.56\pm3.29$ &  $...$  \\
                $\bar{D}^*D^*$ & $10.71\pm1.56$ & $13.60\pm3.14$  &    $51.49\pm7.46$  & $48.53\pm10.98$  & $...$        \\
 \hline \hline
\end{tabular}
\end{table}

The effective  Lagrangians accounting for the interactions between the charmonium states  and a pair of charmed mesons read~\cite{Lin:1999ad,Oh:2000qr,Colangelo:2003sa,Xu:2016kbn,Qian:2023taw}  
\begin{eqnarray}
\mathcal{L}_{\psi DD}&=&ig_{\psi DD} \psi_{\mu}(\partial^{\mu}D\bar{D}-D\partial^{\mu}\bar{D}), \\  
\mathcal{L}_{\psi DD^{\ast}}&=&-g_{\psi DD^{\ast}} \epsilon^{\alpha \beta \mu \nu}\partial_{\alpha} \psi_{\beta}(\partial_{\mu} D_{\nu}^{\ast}\bar{D}+ D \partial_{\mu}\bar{D}^{\ast}_{\nu}), 
\\  
\mathcal{L}_{\psi D^{\ast}D^{\ast}}&=&-ig_{\psi D^{\ast}D^{\ast}} [\psi^{\mu}(\partial_{\mu}D^{\ast}_{\nu}\bar{D}^{\ast\nu}-D^{\ast \nu}\partial_{\mu}\bar{D}^{\ast}_{\nu}) +(\partial_{\mu}\psi_{\nu}D^{\ast\nu}-\psi^{\nu}\partial_{\mu}D^{\ast}_{\nu})\bar{D}^{\ast\mu} \\  \nonumber
&&+ {D}^{\ast\mu} (\psi^{\nu}\partial_{\mu}\bar{D}^{\ast}_{\nu}-\partial_{\mu}\psi_{\nu}\bar{D}^{\ast\nu})], 
\end{eqnarray}
where the $g$ with subscript  are the couplings of the charmonium(like) states  to the charmed mesons.  
There are five vector charmonium(like) states in Table~\ref{Experimentbranching}, which can be categorized into two scenarios.  In scenario I, the $\psi(1S)$ and $\psi(2S)$  as the conventional charmonium states  are below the  $\bar{D}D$ mass threshold, while scenario II  includes the $\psi(3770)$,  $\psi(4040)$, and    $\psi(4160)$ as the charmoniumlike states, all of which are  above the $\bar{D}D$ mass threshold. 
For scenario I, the $\psi(1S)$ and $\psi(2S)$ states are coupled to the $\bar{D}^{(*)}D^{(*)}$ channels. The couplings between $J / \psi$ and $\bar{D}^{(*)}D^{(*)}$ are taken from  the QCD sum rules~\cite{Bracco:2011pg}, as listed in Table~\ref{charmoniacouplings}, while those between $\psi(2S)$ and $\bar{D}^{(*)}D^{(*)}$  are not as straightforward to estimate directly. According to Refs.~\cite{Mehen:2011tp,Guo:2009wr,Li:2011ssa}, the  couplings  between $J / \psi$ and $\bar{D}^{(*)}D^{(*)}$ are  equal to those of $ \psi(2S)$ and $\bar{D}^{(*)}D^{(*)}$. As a result, using the relationship, we obtain the couplings between $ \psi(2S)$ and $\bar{D}^{(*)}D^{(*)}$.   

For scenario II,  the coupling constants  are determined by reproducing the partial decay widths of the charmoniumlike states  decaying into a pair of charmed mesons.  With the partial  width of the decay $\psi(3770)\to \bar{D}D$~\cite{ParticleDataGroup:2024cfk}, the coupling is determined as  $g_{\psi(3770)\bar{D}D} =18.55^{+0.30}_{-0.31}$,  consistent with Refs.~\cite{Zhang:2009gy,Achasov:2021adv}.  Assuming that the total widths of the $\psi(4040)$ and $\psi(4160)$ states are fully saturated by decaying into  a pair of the open charmed mesons~\cite{ParticleDataGroup:2024cfk},   their partial decay widths are estimated to be in Table~\ref{decaywidths}. Then,   we determine the corresponding couplings (denoted by subscript I)  as shown in Table~\ref{charmoniacouplings}. It should be noted that the partial decay widths of $\psi(4040)$ and $\psi(4160)$ vary a lot.  Therefore, we adopt the  partial decay widths of $\psi(4040)$ and $\psi(4160)$ in Refs.~\cite{Husken:2024hmi,Peng:2024blp,Bayar:2019hlw} to determine the couplings  (denoted by subscript II) as shown in Table~\ref{charmoniacouplings}.


\begin{table}[ttt]
\caption{Charmonium(like) states  couplings to a pair of charmed mesons. \label{charmoniacouplings}}
\begin{tabular}{c|ccccccccc}
  \hline\hline
   Couplings  & $g_{\psi(1S)\bar{D}D}$ & $g_{\psi(1S)\bar{D}^*D}$ & $g_{\psi(1S)\bar{D}^*D^*}$

   &  $g_{\psi(2S)\bar{D}D}$ & $g_{\psi(2S)\bar{D}^*D}$ & $g_{\psi(2S)\bar{D}^*D^*}$     \\

   & $5.8\pm0.9$ & $4.0\pm0.6$~GeV$^{-1}$ & $6.2\pm0.9$

    & $5.8^{+1.57}_{-1.39}$ & $4.0^{+1.06}_{-0.94}$~GeV$^{-1}$ & $6.2^{+1.64}_{-1.43}$     \\  \hline

  \multirow{4}{*}{Couplings}  &  $g_{I\psi(4040)\bar{D}D}$& $g_{I\psi(4040)\bar{D}^*D}$ & $g_{I\psi(4040)\bar{D}^*D^*}$  & $g_{\psi(3770)\bar{D}D}$ & $g_{I\psi(4160)\bar{D}^*D}$ & $g_{I\psi(4160)\bar{D}^*D^*}$     \\

   &  $3.1^{+0.22}_{-0.23}$ & $2.47^{+0.17}_{-0.19}$~GeV$^{-1}$ & $3.56^{+0.25}_{-0.27}$  &  $18.55^{+0.30}_{-0.31}$   &    $0.81\pm0.06$ & $1.65\pm{0.12}$    \\
    &  $g_{II\psi(4040)\bar{D}D}$& $g_{II\psi(4040)\bar{D}^*D}$ & $g_{II\psi(4040)\bar{D}^*D^*}$  & $g_{II\psi(4160)\bar{D}D}$ & 
    $g_{II\psi(4160)\bar{D}^*D}$ & $g_{II\psi(4160)\bar{D}^*D^*}$     \\

   &  $5.19^{+0.57}_{-0.64}$ & $2.73^{+0.30}_{-0.34}$~GeV$^{-1}$ & $3.50^{+0.38}_{-0.43}$  &  $3.47^{+0.37}_{-0.42}$   &    $1.18^{+0.13}_{-0.14}$ & $2.02^{+0.22}_{-0.24}$    \\
 \hline \hline
\end{tabular}
\end{table}

Using above Lagrangians, the amplitudes of the triangle diagrams in Fig.~\ref{Fig:loop2}  are written as
\begin{eqnarray}
\mathcal{M}_{a}&=&i^2 \int\frac{d^4 q_3}{(2\pi)^4}\Big[\mathcal{A}_{\mu}(B\to D_{s}^{\ast}\bar{D})\Big]\Big[g_{D^\ast_s DK}p_{1\nu}\Big]\Big[g_{ \psi \bar{D}D}(q_3^{\tau}-q_2^{\tau})\varepsilon_{\tau}(p_2)\Big]  \nonumber  \\  
&&\frac{-g^{\mu\nu}+q^\mu_1 q^\nu_1 /q^2_1}{q^2_1-m^2_1}\frac{1}{q^2_2-m^2_2}\frac{1}{q^2_3-m^2_3}\mathcal{F}(q^2_3,m_3^2),  \\ \nonumber
\mathcal{M}_{b}&=&i^2 \int\frac{d^4 q_3}{(2\pi)^4}\Big[\mathcal{A}_{\mu}(B\to D_{s}\bar{D}^*)\Big]\Big[-g_{D_s D^*K}p_{1\alpha} \Big]   \nonumber\\  &&   \Big[-g_{\psi \bar{D}^*D^*} \varepsilon^{\tau}(p_2) ( g^{\nu\beta} (q_3-q_2)_{\tau} - g^{\beta\tau}(p_2+q_3)_{\nu}+ g^{\tau\nu}(q_2+p_2)_{\beta} )  \Big]    \nonumber\\  && 
\frac{1}{q^2_1-m^2_1}\frac{-g^{\mu\nu}+q^\mu_2 q^\nu_2 /q^2_2}{q^2_2-m^2_2}\frac{-g^{\alpha\beta}+q^\alpha_3 q^\beta_3 /q^2_3}{q^2_3-m^2_3}\mathcal{F}(q^2_3,m_3^2), \\
\mathcal{M}_{c}&=&i^2 \int\frac{d^4 q_3}{(2\pi)^4}\Big[\mathcal{A}_{\omega\mu}(B\to D_{s}^*\bar{D}^*)\Big]\Big[-g_{D_s^* D^*K}\varepsilon_{\rho\phi\eta\alpha}q_1^\rho q_3^{\eta} \Big]  \nonumber\\  &&  \Big[-g_{\psi \bar{D}^*D^*} \varepsilon^{\tau}(p_2) ( g^{\nu\beta} (q_3-q_2)_{\tau} - g^{\beta\tau}(p_2+q_3)_{\nu}+ g^{\tau\nu}(q_2+p_2)_{\beta} )  \Big]    \nonumber    \\  &&
\frac{-g^{\omega\phi}+q_1^{\omega}q_1^{\phi}/q_1^2}{q^2_1-m^2_1}\frac{-g^{\mu\nu}+q^\mu_2 q^\nu_2 /q^2_2}{q^2_2-m^2_2}\frac{-g^{\alpha\beta}+q^\alpha_3 q^\beta_3 /q^2_3}{q^2_3-m^2_3}\mathcal{F}(q^2_3,m_3^2), \\ \nonumber
\mathcal{M}_{d}&=&i^2 \int\frac{d^4 q_3}{(2\pi)^4}\Big[\mathcal{A}_{\omega\mu}(B\to D_{s}^*\bar{D}^*)\Big]\Big[g_{D_s^* D K}p_{1\phi} \Big]\Big[-g_{\psi \bar{D}^*D} \varepsilon^{\tau}(p_2) \varepsilon_{\alpha\tau\beta\nu} p_2^{\alpha} q_2^{\beta} \Big]   \nonumber\\  &&
\frac{-g^{\omega\phi}+q_1^{\omega}q_1^{\phi}/q_1^2}{q^2_1-m^2_1}\frac{-g^{\mu\nu}+q^\mu_2 q^\nu_2 /q^2_2}{q^2_2-m^2_2}\frac{1}{q^2_3-m^2_3}\mathcal{F}(q^2_3,m_3^2), \\
\mathcal{M}_{e}&=&i^2 \int\frac{d^4 q_3}{(2\pi)^4}\Big[\mathcal{A}_{\omega}(B\to D_{s}^*\bar{D})\Big]\Big[-g_{D_s^* D^*K}\varepsilon_{\sigma\phi\lambda\mu} q_1^{\sigma} q_3^{\lambda} \Big]\Big[-g_{\psi \bar{D}^*D} \varepsilon^{\tau}(p_2) \varepsilon_{\alpha\tau\beta\nu} p_2^{\alpha} q_3^{\beta} \Big]   \nonumber\\  &&
\frac{-g^{\omega\phi}+q_1^{\omega}q_1^{\phi}/q_1^2}{q^2_1-m^2_1}\frac{1}{q^2_2-m^2_2}\frac{-g^{\mu\nu}+q^\mu_3 q^\nu_3 /q^2_3}{q^2_3-m^2_3}\mathcal{F}(q^2_3,m_3^2), \\
\mathcal{M}_{f}&=&i^2 \int\frac{d^4 q_3}{(2\pi)^4}\Big[\mathcal{A}(B\to D_{s}\bar{D})\Big]\Big[-g_{D_s D^*K} p_{1\mu} \Big]\Big[-g_{\psi \bar{D}^*D} \varepsilon^{\tau}(p_2) \varepsilon_{\alpha\tau\beta\nu} p_2^{\alpha} q_3^{\beta} \Big]   \nonumber\\  &&
\frac{1}{q^2_1-m^2_1}\frac{1}{q^2_2-m^2_2}\frac{-g^{\mu\nu}+q^\mu_3 q^\nu_3 /q^2_3}{q^2_3-m^2_3}\mathcal{F}(q^2_3,m_3^2),
\end{eqnarray}
where we have introduced a monopole form factor $\mathcal{F}\left(q^{2}, \Lambda^2\right)=\frac{m^2-\Lambda^{2}}{q^{2}-\Lambda^{2}}$. { As demonstrated in Ref.~\cite{Jia:2024pyb},  
the adopted monopole form factor in this work  is consistent with the Pauli-Villars regularization scheme. After introducing the monopole form factor, we can eliminate divergences just like in Pauli-Villars regularization, where the monopole form factor effectively acts as a regulator. Once this “monopole regularization" is applied, the quantities we calculate become finite physical observables with no remaining divergences. In addition, we employ  the monopole form factor to account for the internal structure of the interaction vertices in  the    triangle diagrams,  which is successfully used to construct the hadron-hadron interactions~\cite{Machleidt:1987hj,Tornqvist:1993ng,Liu:2019stu}. This form factor suppresses the high-momentum contributions in the loop integrals, thereby mitigating the unphysical effects arising from large virtual momenta. As a result, the introduction of the form factor not only regularizes the divergent loop integrals but also provides an estimate of the internal structure of the interacting hadrons.      }     The parameter $\Lambda$ can be further parametrized as $\Lambda=m+\alpha\Lambda_{\rm QCD} $ with $\Lambda_{\rm QCD}=0.22 \ {\rm GeV}$~\cite{Cheng:2004ru},  and $m$ is the mass of the exchanged meson.

 With  the amplitudes  for the weak  decays  $B \to \psi K$ given above,
 one can compute their partial decay widths
 \begin{eqnarray}
\Gamma=\frac{1}{2J+1}\frac{1}{8\pi}\frac{|\vec{p}|}{m_{B}^2}{|\overline{M}|}^{2}, \label{eq:FF}
\end{eqnarray}
where $J=0$ is the total angular momentum of the initial $B$ meson, the overline indicates the sum over the polarization vectors of final states, and $|\vec{p}|$ is the momentum of either final state in the rest frame of the $B$ meson.

\section{Results and discussions}
\label{results}

\begin{table}[ttt]
\caption{ Masses and quantum numbers of relevant hadrons needed in this work~\cite{ParticleDataGroup:2024cfk}. \label{mass}}
\begin{tabular}{ccc|cccccc}
  \hline\hline
   Hadron & $I (J^P)$ & M (MeV) &    Hadron & $I (J^P)$ & M (MeV)    \\
  \hline    $K^{0}$ & $\frac{1}{2}(0^-)$ & $497.611$  &    $K^{\pm}$ & $\frac{1}{2}(0^-)$ & $493.677$ \\
$\bar{D}^{0}$ & $1/2(0^-)$ & $1864.84$  &    $D^{-}$ & $1/2(0^-)$ & $1869.66$      \\  
$\bar{D}^{\ast0}$ & $1/2(1^-)$ & $2006.85$ &  $D^{\ast-}$ & $1/2(1^-)$ & $2010.26$   \\
     $D_s^{\pm}$ & $0(0^{-})$ & $1968.35$ &  $D_s^{\ast\pm}$ & $0(1^{-})$ & $2112.20$    \\ 
  $B^{\pm}$ & $\frac{1}{2}(0^-)$ & $5279.41$ &  $B^0$ & $\frac{1}{2}(0^-)$ & $5279.72$  \\
  $\psi(1S)$ & $0(1^-)$ & $3096.9\pm0.006$ &  $\psi(2S)$ & $0(1^-)$ & $3686.097\pm0.011$  \\
  $\psi(4040)$ & $0(1^-)$ & $4039.6\pm4.3$ &  $\psi(3770)$ & $0(1^-)$ & $3773.7\pm0.7$  \\
    $\psi(4160)$ & $0(1^-)$ & $4191\pm5$ &    \\
 \hline \hline
\end{tabular}
\label{tab:masses}
\end{table}

In Table~\ref{mass}, we  tabulate the masses and quantum numbers of relevant particles involved in this work.
     The only  parameter in the present work is the cutoff in the form factor $\mathcal{F}(q^{2}, \Lambda^2)$, which we parametrize through the     dimensionless parameter  $\alpha$.    Since this parameter cannot be determined from first principles, we vary it  from $1$ to $5$ to estimate  the  uncertainties in our calculations, following studies of hadron-hadron interactions~\cite{He:2019rva,Wu:2021udi,Yu:2017zst,Wu:2019rog}.   Moreover, the uncertainties of the couplings (In  Eqs.~(\ref{dddds})-(\ref{dddds1}) and Table~\ref{charmoniacouplings}) in the  vertices of the triangle diagrams lead to uncertainties for the final results, and those of weak interactions lead to around $10\%$ uncertainty\footnote{The uncertainties of the weak vertices in the triangle diagrams are mainly induced by the  $B \to \bar{D}^{(*)}$ transitions.  Using  the experimental  branching fractions of    $\mathcal{B}(B \to \bar{D} \tau \nu_{\tau})=(7.7\pm2.5)\times 10^{-3}$  and $\mathcal{B}(B \to \bar{D}^{*} \tau \nu_{\tau})=(1.88\pm0.20)\%$,  we find that the form factors of $B\to \bar{D}^{(*)}$ in Table~\ref{BtoDformfactor} have $10\%$ uncertainties.  }.    { The short-range contributions are dominated by the Wilson coefficient $a_2$, where the choice of renormalization scale $\mu$ introduces significant variations.  To provide a more robust estimate, we now consider its variation across $\mu \in [0.5 m_b, 1.5 m_b]$, following the prescription of Ref.~\cite{Lu:2009cm}.  }  As a result,   we obtain the uncertainties for the partial decay widths 
originating from the uncertainties of these couplings via a
Monte Carlo sampling within their $1\sigma$ intervals.  In the following, we present the branching fractions  of  the  decays $B \to \psi K$. 



\begin{figure}[ttt]
\centering
\includegraphics[width=7.9cm]{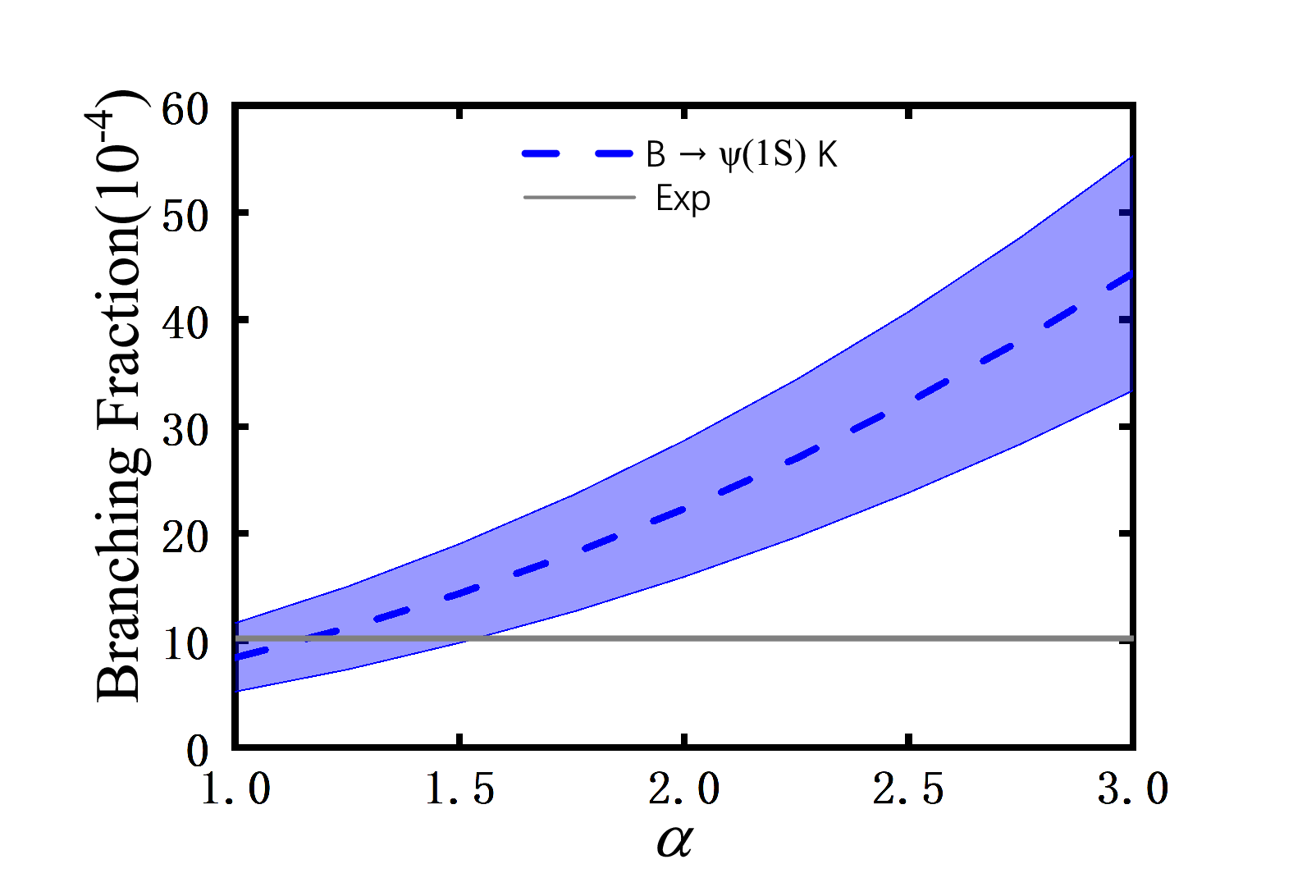}\quad
\includegraphics[width=7.9cm]{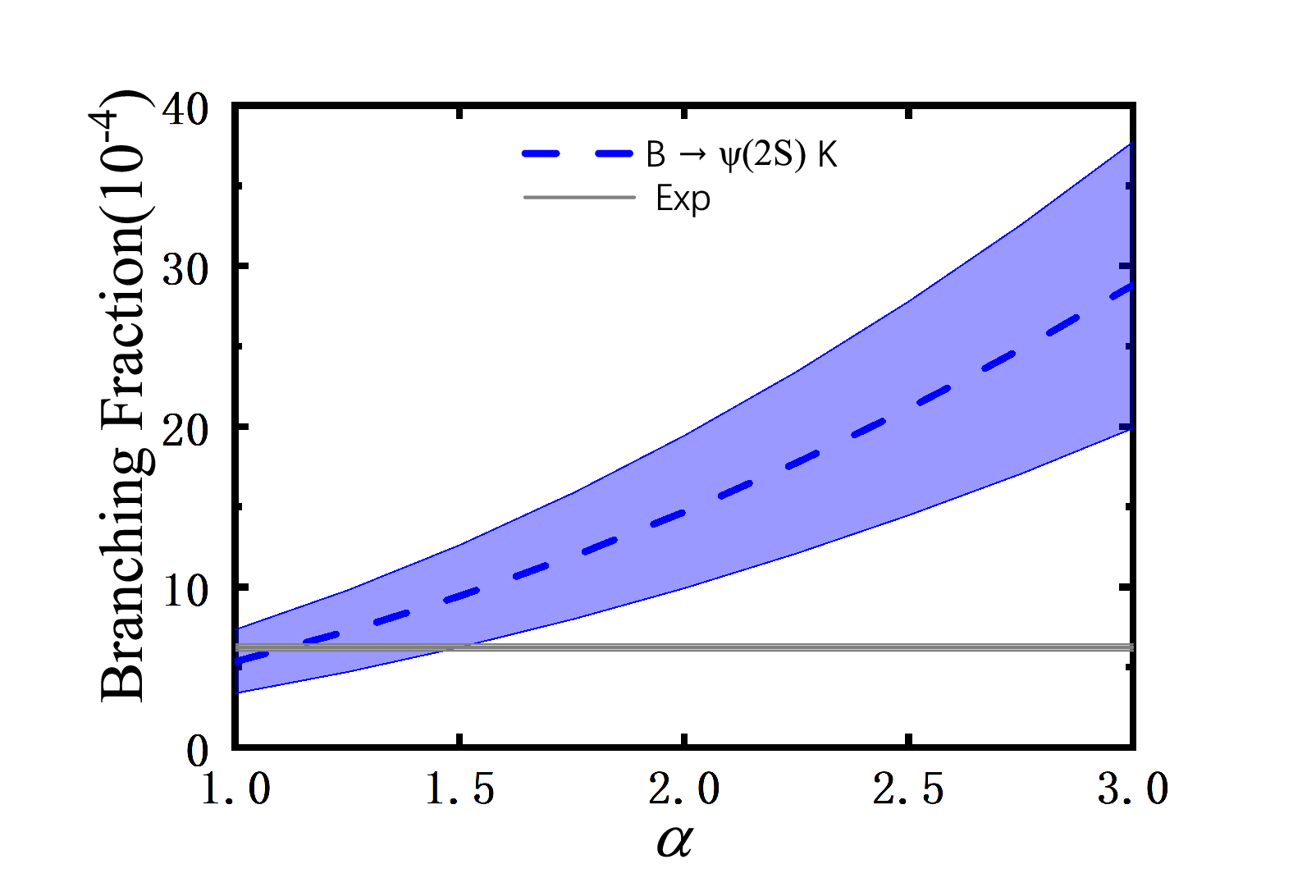}
\caption{ Branching fractions of  the decays $B \rightarrow \psi(1S)K$  and $B \rightarrow   \psi(2S)K$  as the function of $\alpha$. The blue dash  line and band correspond to  the theoretical   central values and uncertainties.  The gray straight  line and band correspond to  the experimental  central values and uncertainties. }  \label{Fig:Jpsi}
\end{figure}

For the states of scenario I,  we present the branching fractions of   the decays of  $B \to \psi(1S) K$ and $B \to \psi(2S) K$ as a function of $\alpha$ in Fig.~\ref{Fig:Jpsi}, where both short-distance and long-distance contributions are taken into account.  We can see that the central values of their  experimental  branching fractions are reproduced when the parameter $\alpha$ is around  $1.2$, which  lies within the reasonable region. Therefore, our model can explain the productions of $\psi(1S)$ and $\psi(2S)$ in $B$ decays.    For a comprehensive analysis of each mechanism's contribution to the branching fractions, we individually present  the    short-distance, long-distance, and total contributions to the decays $B \to   \psi K$   in Table~\ref{branchingvector}. With   only the  short-distance contributions, the branching fractions of the decays of $B \to \psi(1S) K$ and $B \to \psi(2S) K$  are  $ 1.551_{-1.549}^{+1.651}\times 10^{-4}$ and $0.640_{-0.639}^{+0.681}\times 10^{-4}$, respectively,   smaller than the experimental data by one order of magnitude, which implies that the nonfactorization effects of these decays are significant. With only long-distance contributions, the branching fractions of the decays of $B \to \psi(1S) K$ and $B \to \psi(2S) K$ become  larger than those with only short-distance contributions. Once taking into account both short-distance and long-distance contributions, i.e., total  contributions,  the branching fractions of the decays of  $B \to \psi(1S) K$ and $B \to \psi(2S) K$  are enhanced since the amplitudes of tree diagrams  and triangle diagrams add constructively. From the above analysis,    it is demonstrated  that  the  FSIs play a crucial role in the decays  of $B \to \psi(1S) K$ and $B \to \psi(2S) K$.

\begin{figure}[ttt]
\centering
\includegraphics[width=5.4cm]{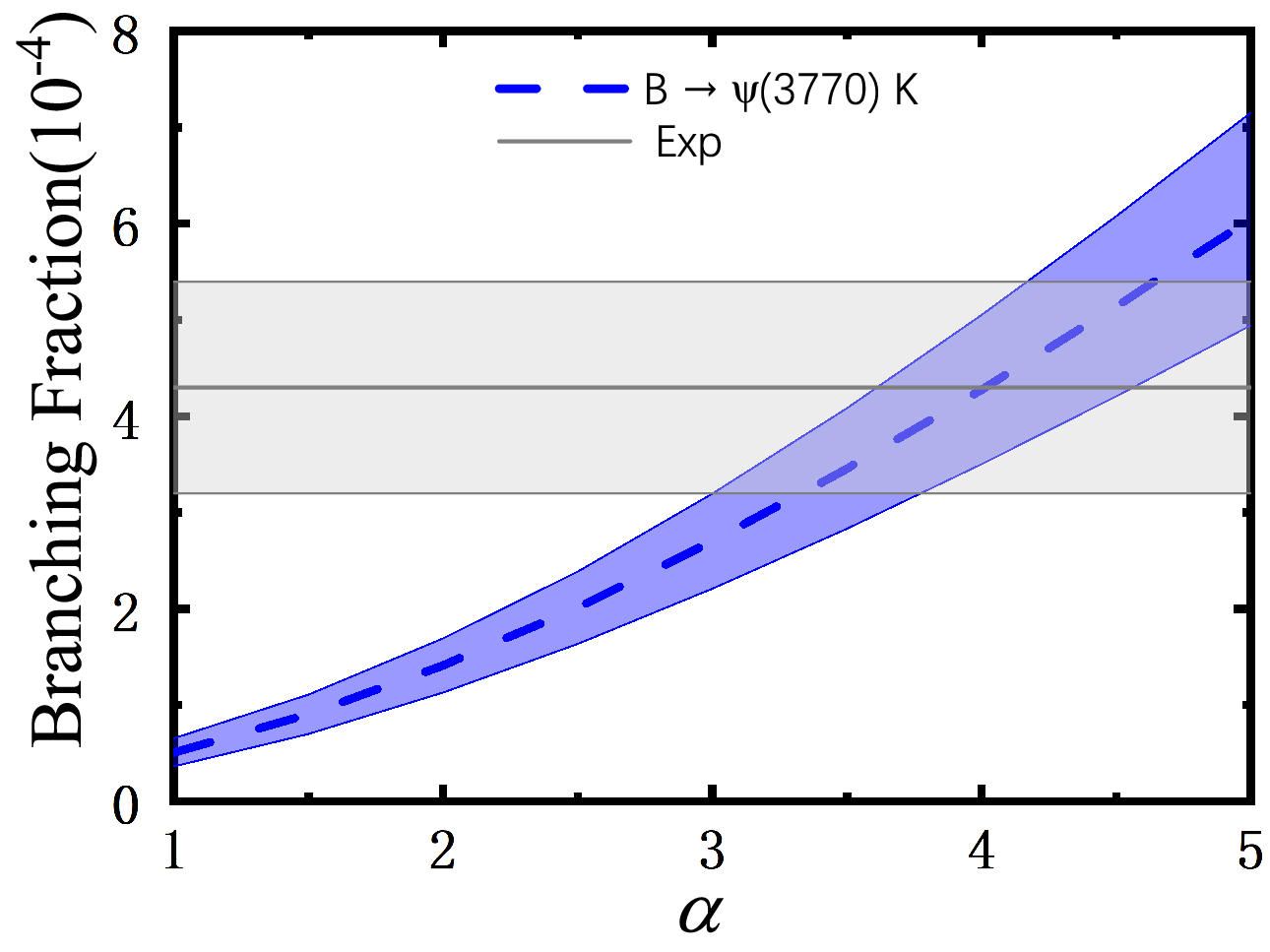}
\includegraphics[width=5.4cm]{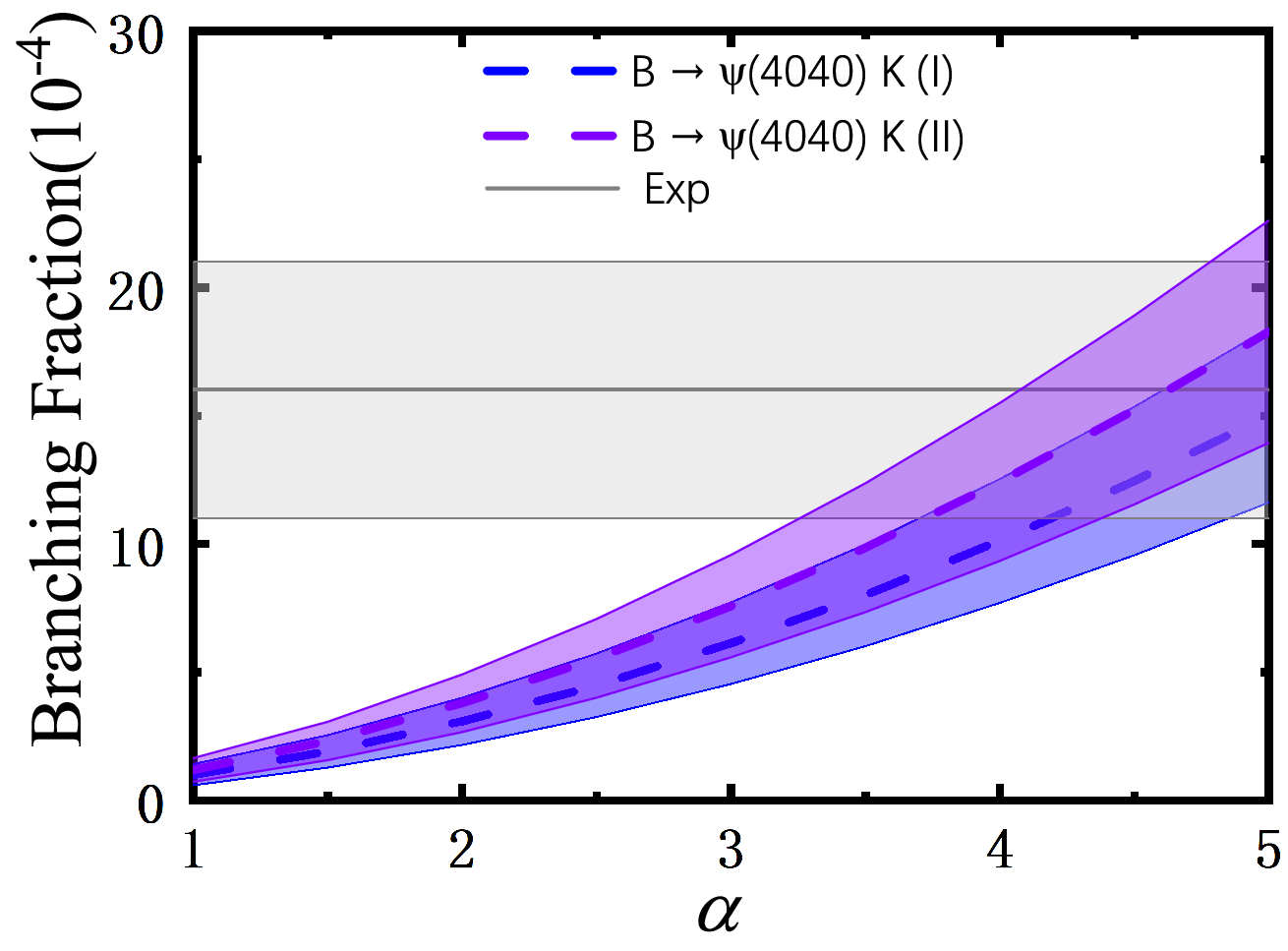}
\includegraphics[width=5.4cm]{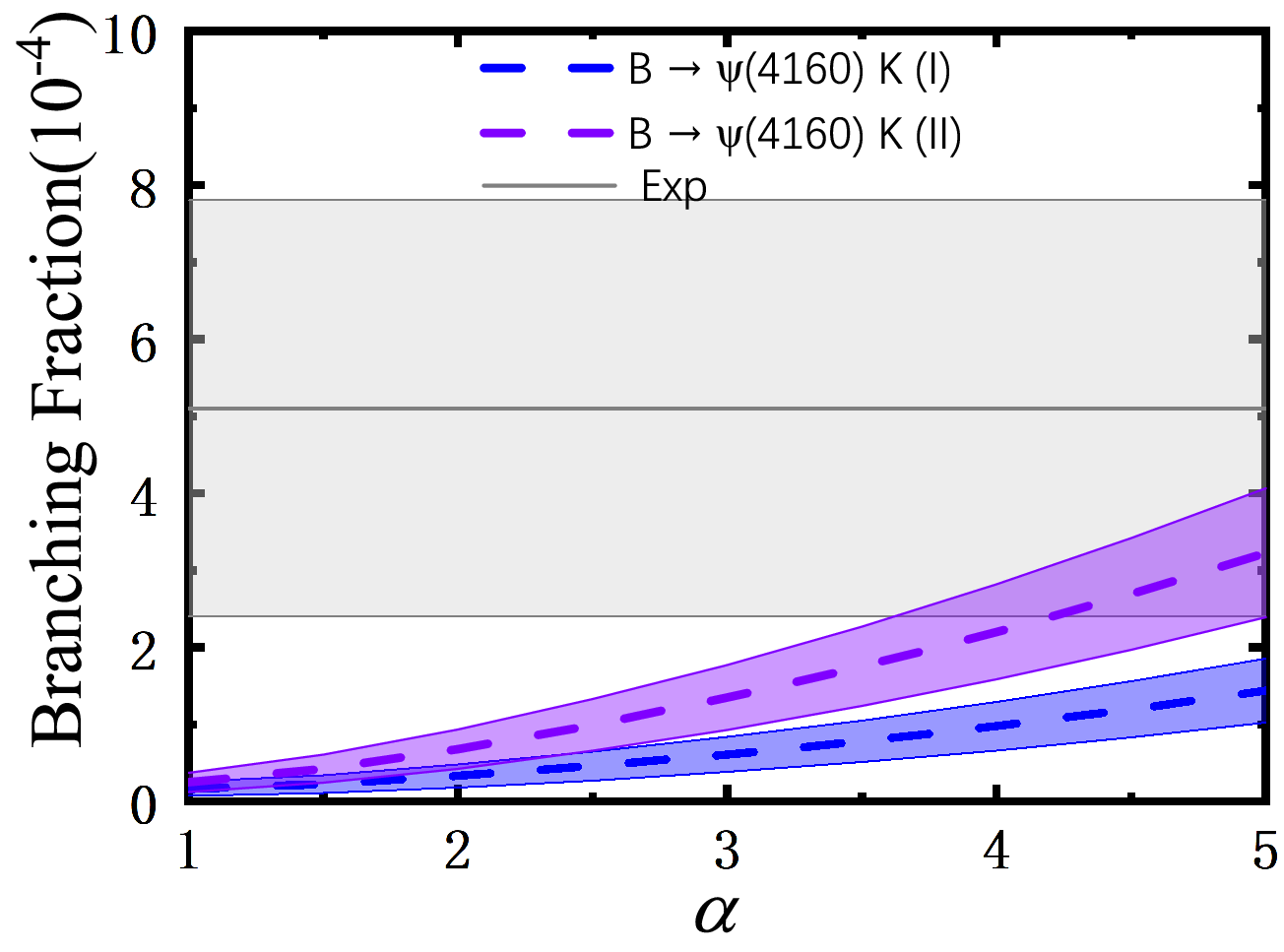}
\caption{Branching fractions of $B \rightarrow \psi(3770)K$, $B \rightarrow \psi(4040)K$, and $B \rightarrow \psi(4160)K$  as the function of $\alpha$. The blue dash  line and band correspond to  the theoretical   central values and uncertainties.  The gray straight  line and band correspond to  the experimental  central values and uncertainties. The purple line represents the results obtained by using the masses and decay widths retrieved from Refs.~\cite{Husken:2024hmi,Peng:2024blp,Bayar:2019hlw} as input for the calculations. } \label{fig:vector charmonium}
\end{figure}

For the states of scenario II,   we   present the branching fractions of   $B \to \psi(3770) K$, $B\to \psi(4040) K$, and $B\to \psi(4160) K$ decays  as a function of $\alpha$ in Fig.~\ref{fig:vector charmonium}. Different from those of scenario I,  larger values of the parameter $\alpha$ are  necessary to explain the experimental data.  Using the couplings $g_{\psi \bar{D}^{(*)}D^{(*)}}$ from  Ref.~\cite{ParticleDataGroup:2024cfk}, we find  the parameter $\alpha $ to be  around $4$, $5$, and $11.5$ for  the     decays of  $B \to \psi(3770) K$, $B \to \psi(4040) K$, and $B \to \psi(4160) K$, respectively. These values correspond to  the blue line and band in Fig.~\ref{fig:vector charmonium}.   In contrast to  the $\psi(3770)$ state, the  widths  of  the  $\psi(4040)  $ and $\psi(4160)$  states  are still under debate. Using the results from  Refs.~\cite{Husken:2024hmi,Peng:2024blp}  as inputs, we  recalculate the branching fractions for    $B \to \psi(4040) K$ and     $B \to \psi(4160) K$ decays, shown as the purple line and band  in Fig.~\ref{fig:vector charmonium}. Notably,    the parameter $\alpha $  decreases to approximately $4.5$ and $6.5$ for the $B  \to \psi(4040) K $ and $B  \to \psi(4160) K $ decays.   Therefore, the productions of  $\psi(3770)$ and  $\psi(4040)$ in $B$ decays can be  explained in our model.   However, the parameter $\alpha$ for the decay $B  \to \psi(4160) K $ remains unnatural in our model\footnote{Assuming  $\psi(4160)$ and $Y(4230)$ as same states,   $\psi(4160)$ would have strong coupling to $\bar{D}D_1$,  implying  that  weak vertices $B \to \bar{D}^{(*)}D_1$ combing the FSIs~\cite{Liu:2024hba} would contribute to the branching fractions of the decay $B  \to \psi(4160) K $ and change the parameter $\alpha$ in the reasonable region. }, which also supports  the exotic nature of $\psi(4160)$ from the perspective of $B$ decays, similar to the conclusions obtained from analyzing the invariant mass distributions~\cite{Zhou:2023yjv}.

Similar to the states of scenario I,   we separately list  the    short-distance, long-distance, and total contributions to the decays of  $B \to \psi(3770) K$, $B \to \psi(4040) K$, and $B \to \psi(4160) K$   in Table~\ref{branchingvector}.  The short-distance contributions alone yield  branching fractions of $\mathcal{B}[B \to \psi(3770) K]=7.0_{-7.0}^{+7.5}\times10^{-6}$, $\mathcal{B}[B \to \psi(4160) K]=9.0_{-9.0}^{+9.6}\times10^{-6}$, and $\mathcal{B}[B \to \psi(4040) K]=1.93_{-1.93}^{+2.05}\times10^{-5}$,  which are nearly two orders of magnitude smaller than the experimental values. This significant discrepancy indicates the importance of nonfactorizable effects in these decay processes.   In contrast, the long-distance contributions dominant the decay amplitudes, producing substantially larger branching fractions. When incorporating both short-distance and long-distance contributions coherently, we find the enhancement in the branching fractions due to constructive interference between the two amplitudes.  However, the interference effect is not as prominent, indicating that FSIs play a dominant role in the processes of the $B \to \psi(3770) K$, $B \to \psi(4040) K$, and $B \to \psi(4160) K$   decays.

As indicated in~\cite{Jia:2024pyb}, the ratio of  $C$-type topology diagram   to  $T$-type topology diagram is estimated to be $\frac{|C|}{|T|}\sim \mathcal{O}\frac{\Lambda_{QCD}^{hadron}}{m_{Q}}$.  This ratio is approximately unity in charm decays but reduces to $1/4$ in bottom decays, demonstrating less pronounced color suppressed effects in bottom hadron decays compared to charm hadron decays, which is identified by the  experimental branching fraction measurements.  Nevertheless, the nonfactorization effects in bottomed hadron decays are still sizable~\cite{Wang:2024oyi,Duan:2024zjv,Geng:2025yna}. 
Our results in Table~\ref{branchingvector} reveal that the  nonfactorization effects  are significant in the decays of the $B$ meson into the vector charmonium(like) states and a $K$ meson.   This provides compelling evidence for the dominant role of FSIs in these processes. In particular, the nonfactorization effects in  the  $B \to \psi(3770) K$, $B \to \psi(4040) K$, and $B \to \psi(4160) K$ decays  are more prominent than     
those in   $B \to \psi(1S) K$ and $B \to \psi(2S) K$ decays.  

\begin{table}[ttt]
\caption{Branching fractions~($10^{-4}$) of $B$ meson decaying into vector charmonium(like) states and kaon meson. \label{branchingvector}}
\begin{tabular}{c|ccccccccc}
  \hline\hline
   Decay modes &  Short-distance  & Long-distance &  Total   & Experiments   \\  \hline
       $\to \psi(1S)$ & {$1.551_{-1.549}^{+1.651}$} & $2.818\sim89.519$ & $8.402\sim112.177$   &  $10.20\pm0.19$   \\ \hline
          $\to \psi(2S)$ & {$0.640_{-0.639}^{+0.681}$} & $2.299\sim59.555$ & $5.351\sim71.948$   &  $6.24\pm0.21$  \\  \hline
      \multirow{2}{*}{$\to \psi(4040)$} & \multirow{2}{*}{{$0.193_{-0.193}^{+0.205}$}} & $0.331\sim11.847$ & $1.006\sim 15.019$  &  \multirow{2}{*}{$16.0\pm 5.0$}   \\  
  &  & $0.432\sim14.809$ & $1.185\sim 18.267$  &     \\   \hline  
    $\to \psi(3770)$ & {$0.070_{-0.070}^{+0.075}$} & $0.269\sim5.954$ & $0.508\sim6.048$   &  $4.3\pm1.1$   \\   \hline   
        \multirow{2}{*}{$\to \psi(4160)$} & \multirow{2}{*}{{$0.090_{-0.090}^{+0.096}$}} & $0.010\sim0.809$ & $0.161\sim1.437$  &  \multirow{2}{*}{$5.1\pm 2.7$}     \\   
          &  & $0.035\sim2.216$ & $0.247\sim3.228$  &       \\
 \hline \hline
\end{tabular}
\label{tab:masses}
\end{table}

\section{Summary and Discussion}
\label{sum}

The decays of the $B$ meson into a series of  vector charmonium(like) states and a $K$ meson have been observed, providing new opportunities to investigate the nature of charmonium(like) states and their production mechanisms in   $B$ decays.   
 In this work, we  focused  on the decays   $B \to \psi K$ [where $\psi$ denotes $\psi(1S)$, $\psi(2S)$, $\psi(4040)$, $\psi(3770)$, and $\psi(4160)$],  for which the   branching fractions have been experimentally measured.  We  analyzed both  short-distance and long-distance contributions to  these decays,  illustrated by the tree diagrams and triangle diagrams, respectively.  The short-distance contributions are evaluated within the framework of naive factorization, whereas the long-distance contributions are investigated through  FSIs. Using the effective Lagrangian approach, we systematically calculated both the short-distance and long-distance contributions to the  $B \to \psi K$ decays.  

In our model, the experimental  branching fractions of  $B \to \psi(1S) K$ and   $B \to \psi(2S) K$ decays  are reproduced with $\alpha \approx 1$,  while those of $B \to \psi(3770) K$ and  $B \to \psi(4040) K$ decays require  $\alpha \approx 4$. However,  the  branching fraction of the decay $B \to \psi(4160) K$ can only be accommodated with an unnaturally large  $\alpha$, implying  the exotic nature of $ \psi(4160)$ from the perspective of $B$ decays. The dominant uncertainty in our model arises from the couplings between the charmonium(like) states  and  $\bar{D}^{(*)}D^{(*)}$, which lead to the uncertainties of our results. However, the decay constants of these vector charmonium(like) states are determined directly from their electronic decay widths, ensuring that the results calculated by only  short-distance contributions remain robust.  Since results within only  short-distance contributions play minor role in these decays, we infer that the role of  long-distance contributions to these decays are more prominent. Therefore, we conclude that  the FSIs play the dominant role in    $B \to \psi K$ decays. Our findings justify studying the production of    charmoniumlike states  in $B$ decays via FSIs, reinforcing the importance of this mechanism in understanding such charmoniumlike states.

\section{Acknowledgments}
 
We are  grateful to Li-Sheng Geng,  Qin Chang, Fu-Sheng Yu, Li-Ting Wang, and  Zhu-Ding Duan  for useful discussions.  
 This work is supported by the National Natural Science Foundation of China under Grants Nos.12105007 and 12405093, as well as supported, in part, by National Key Research and Development Program under Grant No.2024YFA1610504.

\appendix

\bibliography{biblio}
\end{document}